\documentclass[twocolumn,twoside,aps,prc,nofootinbib]{revtex4}

\usepackage{graphicx}

\usepackage{isotope}
\usepackage{physics}

\usepackage{url}
\usepackage{hyperref} 
\usepackage{orcidlink}

\usepackage[normalem]{ulem}

\newcommand{\sigmabf}{\mbox{\boldmath $\sigma$}}

\def\SCNTHuizhou{$^{(1)}$Southern Center for Nuclear-Science Theory (SCNT), Institute of Modern Physics, Chinese Academy of Sciences, Huizhou 516000, China}
\def\KinrKyiv{$^{(2)}$Institute for Nuclear Research, National Academy of Sciences of Ukraine, Kyiv 03680, Ukraine}
\def\HISTKLLanzhou{$^{(3)}$Heavy Ion Science and Technology Key Laboratory, Institute of Modern Physics, Chinese Academy of Sciences, Lanzhou 730000, China}
\def\UnivCASBeijing{$^{(4)}$School of Nuclear Sciences and Technology, University of Chinese Academy of Sciences, Beijing 101408, China}
\def\sunyatsen{$^{(5)}$School of Physics and Astronomy, Sun Yat-Sen University, Zhuhai 519082, China}
\def\sunyatsenlp{$^{(6)}$Sino-French Institute of Nuclear Engineering and Technology, Sun Yat-Sen University, Zhuhai 519082, China}

\begin{document}

\title{Quantum dominance of coherent bremsstrahlung in $\isotope[124]{Sn} + \isotope[124]{Sn}$ scattering at 25 MeV/u}

\author{Sergei~P.~Maydanyuk$^{(1,2)}$\,\orcidlink{0000-0001-7798-1271}}\email{sergei.maydanyuk@impcas.ac.cn}
\author{Ju-Jun Xie$^{(1,3,4)}$\,\orcidlink{0000-0001-9888-5924}}\email{xiejujun@impcas.ac.cn}
\author{Peng-Ming~Zhang$^{(5)}$\,\orcidlink{0000-0002-1737-3845}}\email{zhangpm5@mail.sysu.edu.cn}
\author{Li-Ping~Zou$^{(6)}$\,\orcidlink{0000-0001-8976-9171}}\email{zoulp5@mail.sysu.edu.cn}

\affiliation{\SCNTHuizhou}
\affiliation{\KinrKyiv}
\affiliation{\HISTKLLanzhou}
\affiliation{\UnivCASBeijing}
\affiliation{\sunyatsen}
\affiliation{\sunyatsenlp}

\date{\small\today}

\begin{abstract}

We present 
quantum-mechanical calculations of bremsstrahlung 
in the $^{124}$Sn+$^{124}$Sn at 25 MeV/u reproducing the measured photon spectrum over the full energy range.
For the first time, we quantitatively determine the incoherent-to-coherent ratio in the photon spectrum. This ratio is extremely small, ranging from $10^{-11}$ to $10^{-4}$, which demonstrates that coherent emission dominates throughout the measured energy range. This behavior is in sharp contrast to proton--nucleus scattering, where incoherent emission dominates because of the leading role of nucleon magnetic moments. This contrast is illustrated by the TAPS Collaboration data for $p + ^{197}{\rm Au}$ collisions at a proton beam energy of 190~MeV, where the corresponding ratio reaches $10^{3}$--$10^{5}$. 
We find that incoherent-to-coherent ratios explain 
the difference between the two spectra
in unified picture:
(1) In proton--nucleus scattering, the spectrum contains a pronounced hump,
(2) In $\isotope[124]{Sn} + \isotope[124]{Sn}$ scattering, the spectrum decreases monotonically and has a nearly logarithmic shape.
%
%
Our results identify a previously unexplored quantum regime of bremsstrahlung emission in nuclear reactions and open a new route for studying coherent effects in heavy-ion collisions.

\end{abstract}



\maketitle

\vspace{-0.5mm}
\emph{Introduction.}\;
Bremsstrahlung photons produced in heavy-ion reactions provide a clean probe of many-nucleon dynamics in nuclear collisions. As the nucleons interact during the collision, their electric charges and magnetic moments generate electromagnetic radiation that can be measured experimentally. Although the full many-body dynamics remains difficult to solve, the separation of coherent and incoherent bremsstrahlung components can provide direct information on the underlying reaction mechanisms. A reliable use of this observable therefore requires a unified theoretical framework in which both emission mechanisms are treated consistently.

The ratio of incoherent to coherent bremsstrahlung is highly sensitive to the structure of the nuclei participating in the reaction. This sensitivity has been used to extract structural information for $\alpha$ particles~\cite{Maydanyuk_Zhang_Zou.2016.PRC}, deuterons~\cite{Shaulskyi_Maydanyuk_Vasilevsky.2024.PRC.c110}, and a series of nuclei in reactions~\cite{Maydanyuk_Zhang_Zou.2016.PRC}. The same ratio strongly affects the shape of the bremsstrahlung cross section. High-precision measurements of bremsstrahlung cross sections or probabilities have been reported for proton--nucleus scattering~\cite{Goethem.2002.PRL}, $\alpha$ decay~\cite{Boie.2007.PRL,Boie.2009.PhD}, spontaneous fission~\cite{Eremin.2010.IJMPE,Maydanyuk.2010.PRC}, and heavy-ion collisions~\cite{Qin:2024plb,br124Sn124Snexp.2026.PRC,Xu:2025prr}.

Bremsstrahlung spectra can also provide information on the nuclear part of the interaction potential in proton--nucleus and pion--nucleus reactions; examples include $\pi^{+}+\isotope[44]{Ca}$ at $E_{\pi}=116$~MeV and $p+\isotope[208]{Pb}$ and $p+\isotope[197]{Au}$ at proton energies of 140 and 190~MeV 
\cite{Maydanyuk_Zhang_Zou.2018.PRC}. Such information can be extracted from measured bremsstrahlung data only if coherent and incoherent emission are described within a common framework.

Despite early recognition of this potential~\cite{Maydanyuk_Zhang_Zou.2016.PRC,Shaulskyi_Maydanyuk_Vasilevsky.2024.PRC.c110}, the use of bremsstrahlung as a tool for nuclear-structure studies in reactions remains limited. Important progress has been made for light-ion reactions using microscopic cluster models~\cite{1985NuPhA.443..302B,1990PhRvC..41.1401L,1990PhRvC..42.1895L,1991NuPhA.529..467B,1991PhRvC..44.1695L,1992NuPhA.550..250B,Liu.1992.FBS,1993nuco.conf..423K,Dohet-Eraly.2011.JPCS,Dohet-Eraly.2011.PRC,Dohet-Eraly.2013.PRC,2013JPhCS.436a2030D,Dohet-Eraly.2013.PhD,2014PhRvC..89b4617D,2014PhRvC..90c4611D,Timofeyuk.2024.PRC.v110}. However, a broad set of high-precision data, including proton--nucleus scattering~\cite{Goethem.2002.PRL,Edington.1966.NP,Koehler.1967.PRL,Kwato_Njock.1988.PLB,Pinston.1989.PLB,Pinston.1990.PLB}, $\alpha$ decay of $\isotope[210]{Po}$~\cite{Boie.2007.PRL,Boie.2009.PhD,DArrigo.1994.PHLTA,Kasagi.1997.JPHGB,Kasagi.1997.PRLTA,Maydanyuk.2008.EPJA,Giardina.2008.MPLA}, and spontaneous fission of $\isotope[252]{Cf}$~\cite{Eremin.2010.IJMPE,Maydanyuk.2010.PRC,Ploeg.1995.PRC,Kasagi.1989.JPSJ,Luke.1991.PRC,Varlachev.2007.BRASP,Hofman.1993.PRC,Pandit.2010.PLB}, has not yet been systematically analyzed within such a framework. Measurements of $\gamma$ emission in fusion--evaporation reactions, such as $^{32}$S+$^{100}$Mo and $^{36}$S+$^{92}$Mo at beam energies of 196 and 214.2~MeV, have been used to study fusion dynamics~\cite{Pierroutsakou.2005.PRC}. Gamma decays in the pygmy-dipole region of heavy nuclei have also been investigated~\cite{Kmiecik.Maj.2020.ActaPP}. Nevertheless, systematic theoretical studies of bremsstrahlung emission in intermediate- and high-energy heavy-ion collisions remain scarce. Recent measurements~\cite{Qin:2024plb} have therefore promising interest in this problem.

In reactions involving heavy nuclei, a central issue is the relative strength of incoherent and coherent bremsstrahlung. 
Incoherent emission originates from the dynamics of individual nucleons, whereas coherent emission is associated with the collective two-body motion of the colliding nuclei 
(we follow Ref.~\cite{Maydanyuk_Vasilevsky.2023.PRC} for these definitions). The relative importance of these two components differs substantially between proton--nucleus and nucleus--nucleus reactions. In proton--nucleus scattering, incoherent emission is known to dominate~\cite{Goethem.2002.PRL,Maydanyuk.2023.PRC.delta,Maydanyuk_Zhang.2015.PRC}. For heavy-ion reactions, however, the incoherent-to-coherent ratio has not yet been derived within a rigorous quantum-mechanical framework capable of treating both processes simultaneously. This motivates the central question of the present work: how large are the incoherent and coherent contributions in heavy-nucleus collisions, and how can they be determined from measured photon spectra?

We address this question by extending a formalism previously developed for proton--nucleus scattering~\cite{Maydanyuk.2023.PRC.delta,Maydanyuk_Vasilevsky.2023.PRC,Maydanyuk_Zhang.2015.PRC} to heavy-ion collisions. Because different definitions of coherent and incoherent bremsstrahlung exist in the literature, we adopt the formulation introduced in Ref.~\cite{Maydanyuk_Vasilevsky.2023.PRC}. We perform quantum-mechanical calculations of bremsstrahlung emission in heavy-ion reactions and, by reproducing the experimental spectra, determine for the first time the coherent and incoherent contributions. The results reveal a striking dominance of coherent emission, a quantum effect that challenges existing expectations and provides a new way to explore collective dynamics in heavy-ion collisions.

\vspace{1.5mm}
\emph{Model.}\;
For the description of bremsstrahlung photon emission, we start from the formalism developed for proton--nucleus scattering~\cite{Maydanyuk.2023.PRC.delta}, which has been tested in a number of nuclear-physics applications~\cite{Maydanyuk.2012.PRC,Maydanyuk_Zhang.2015.PRC,Maydanyuk_Zhang_Zou.2016.PRC,Liu_Maydanyuk_Zhang_Liu.2019.PRC.hypernuclei,Maydanyuk_Zhang_Zou.2018.PRC,Maydanyuk_Zhang_Zou.2019.PRC.microscopy}. For nucleus--nucleus scattering, this formalism must be generalized to include a many-nucleon projectile and a many-nucleon target. In the laboratory frame, the Hamiltonian of a system with photon emission is written as the many-nucleon generalization of the Pauli Hamiltonian, $\hat{H}=\hat{H}_{0}+\hat{H}_{\gamma}$, where
\begin{equation}
\begin{array}{lllll}
\vspace{1mm}
  \hat{H}_{0} & = &
  \displaystyle\sum_{i=1}^{A} 
  \displaystyle\frac{\vu{p}_{Ai}^{2}}{2m_{Ai}} +
  \displaystyle\sum_{j=1}^{B} 
  \displaystyle\frac{\vu{p}_{Bj}^{2}}{2m_{Bj}} +
  V(\vb{r}_{A1} \ldots \vb{r}_{AA},  \vb{r}_{B1} \ldots \vb{r}_{BB}), \\

\vspace{1mm}
  \hat{H}_{\gamma} & = &
  \displaystyle\sum_{i=1}^{A}
  \biggl\{
    - \displaystyle\frac{z_{Ai} e}{m_{Ai}c}\; \vu{p}_{Ai} \cdot \vb{A}_{Ai} +
    \displaystyle\frac{z_{Ai}^{2}e^{2}}{2m_{Ai}c^{2}} \vb{A}_{Ai}^{2} - \\
  & - &
    \displaystyle\frac{z_{Ai}e\hbar}{2m_{Ai}c}\, \sigmabf \cdot \vb{rot A}_{Ai} +
    z_{Ai}e\, A_{Ai,0}
  \biggr\}\; + \\

  & + &
  \displaystyle\sum_{j=1}^{B}
  \biggl\{
    - \displaystyle\frac{z_{Bj} e}{m_{Bj}c}\; \vu{p}_{Bj} \cdot \vb{A}_{Bj} +
    \displaystyle\frac{z_{Bj}^{2}e^{2}}{2m_{Bj}c^{2}} \vb{A}_{Bj}^{2} - \\
  & - &
  \displaystyle\frac{z_{Bj}e\hbar}{2m_{Bj}c}\, \sigmabf \cdot \vb{rot A}_{Bj} +
    z_{Bj}e\, A_{Bj,0}
  \biggr\}.
\end{array}
\label{eq.pauli.5}
\end{equation}
Here, $\hat{H}_{0}$ is the Hamiltonian governing the evolution of the nucleons in the two nuclei in the absence of photon emission, and $\hat{H}_{\gamma}$ is the operator describing bremsstrahlung emission. The quantities $z_{Ai}$ and $z_{Bj}$ are the electric charges of nucleon $i$ in nucleus $A$ and nucleon $j$ in nucleus $B$, respectively. The quantities $m_{Ai}$ and $m_{Bj}$ are the corresponding nucleon masses, $\vu{p}_{i}=-i\hbar\,\vb{d}/\vb{dr}_{i}$ is the momentum operator of nucleon $i$, and $V(\vb{r}_{A1}\ldots\vb{r}_{AA},\vb{r}_{B1}\ldots\vb{r}_{BB})$ is the general interaction potential between the nucleons of the two nuclei. The vector $\sigmabf$ denotes the Pauli matrices, $A_{i}=(\vb{A}_{i},A_{i,0})$ is the electromagnetic four-potential at the position of nucleon $i$, and $A$ and $B$ are the mass numbers of the two nuclei.

We neglect the terms proportional to $\vb{A}_{j}^{2}$ and $A_{j,0}$ and use the Coulomb gauge. The magnetic moments of the nucleons are included following Ref.~\cite{Maydanyuk.2023.PRC.delta}. Specifically, 
the Dirac magnetic moment $\mu_{i}^{\rm (Dir)}=z_{i}e\hbar/(2m_{i}c)$ of nucleon $i$ is replaced by $\mu_{i}^{\rm (an)}\mu_{N}$, where $\mu_{N}=e\hbar/(2m_{\rm p}c)=3.152 451 2550\;10^{-14}$~MeV T$^{-1}$ is the nuclear magneton, $\mu_{\rm p}^{\rm (an)}=2.79284734462$ is the anomalous magnetic moment of the proton, and $\mu_{\rm n}^{\rm (an)}=-1.91304273$ is the anomalous magnetic moment of the neutron, both measured in units of $\mu_N$~\cite{RewPartPhys_PDG.2018}.

To study photon emission, we introduce relative coordinates for the electric charges and magnetic moments of the nucleons in nucleus--nucleus scattering and rewrite the formalism in terms of these coordinates and the corresponding relative momenta [see Sec.~I.A of the Supplemental Material~\cite{thispaper.2026.Supplementary}]. In the laboratory frame, the photon-emission operator can then be written as $\hat{H}_{\gamma}=\hat{H}_{P}+\hat{H}_{p}+\Delta\hat{H}_{\gamma E}+\Delta\hat{H}_{\gamma M}+\hat{H}_{k}$. Here, $\hat{H}_{P}$ is associated with the motion of the full nucleus--nucleus system, $\hat{H}_{p}$ describes coherent emission, $\Delta\hat{H}_{\gamma E}$ and $\Delta\hat{H}_{\gamma M}$ describe incoherent emission of electric and magnetic type, and $\hat{H}_{k}$ is the background term. These terms are derived in Eqs.~(S4)--(S9) of the Supplemental Material~\cite{thispaper.2026.Supplementary}.

The photon-emission matrix element $\langle\Psi_{f}|\,\hat{H}_{\gamma}\,|\Psi_{i}\rangle$ is defined using the wave functions $\Psi_i$ and $\Psi_f$ of the full nuclear system before photon emission (the initial state) and after photon emission (the final state), respectively. After the transformations described above, the full bremsstrahlung matrix element can be written as
$\langle \Psi_{f} |\, \hat{H}_{\gamma} |\, \Psi_{i} \rangle=\sqrt{2\pi c^{2}/(\hbar w_{\rm ph})}\,M_{\rm full}$, where $M_{\rm full}=M_{P}+M_{p}^{(E)}+M_{p}^{(M)}+M_{k}+M_{\Delta E}+M_{\Delta M}$. Here, $M_{p}^{(E)}$ and $M_{p}^{(M)}$ are the electric and magnetic coherent terms, $M_{\Delta M}$ and $M_{k}$ are the magnetic incoherent and background terms, and $M_{P}$ is associated with the center-of-mass motion of the nucleus--nucleus system [see Eqs.~(S12)--(S17) of the Supplemental Material~\cite{thispaper.2026.Supplementary}]:

\vspace{-5.5mm}
\begin{widetext}
\begin{equation}
\begin{array}{lcl}
\vspace{-1.0mm}
  M_{P} & = &
  \hbar\, (2\pi)^{3}\,
  \displaystyle\frac{\mu_{N}}{m_{A} + m_{B}}\,
  \displaystyle\sum\limits_{\alpha=1,2}
  \displaystyle\int\limits_{}^{}
    \Phi_{A - B, f}^{*} (\vb{r})\;

  \biggl\{
    2\, m_{\rm p}\;
    \Bigl[
      e^{-i\, c_{B}\, \vb{k_{\rm ph}} \vb{r}} D_{A,P\, {\rm el}} +
      e^{ i\, c_{A}\, \vb{k_{\rm ph}} \vb{r}} D_{B,P\, {\rm el}}
    \Bigr]\, \vb{e}^{(\alpha)} \cdot \vb{K}_{i} + \\

  & + &
  i\,
  \Bigl[
    m_{A}\, e^{-i\, c_{B}\, \vb{k_{\rm ph}} \vb{r}}\, \vb{F}_{A,P\, {\rm mag}} +
    m_{B}\, e^{ i\, c_{A}\, \vb{k_{\rm ph}} \vb{r}}\, \vb{F}_{B,P\, {\rm mag}}
  \Bigr]\, \cdot
    \bigl[ \vb{K}_{i} \cp \vb{e}^{(\alpha)} \bigr]
  \biggr\} \cdot
  \Phi_{A - B, i} (\vb{r})\; \vb{dr},
\end{array}
\label{eq.13.1.1}
\end{equation}

\vspace{-5.0mm}
\begin{equation}
\begin{array}{lll}
\vspace{-0.5mm}
  M_{p}^{(E)} & = &
  i \hbar\, (2\pi)^{3}\, \displaystyle\frac{2\, \mu_{N}\,  m_{\rm p}}{\mu}\;
  \displaystyle\sum\limits_{\alpha=1,2}
    \vb{e}^{(\alpha)}
  \displaystyle\int\limits_{}^{}
    \Phi_{A - B, f}^{*} (\vb{r})\;
    Z_{\rm eff} (\vb{k}_{\rm ph}, \vb{r}) \,
    e^{-i\, \vb{k}_{\rm ph} \vb{r}}\;
    \vb{\displaystyle\frac{d}{dr}}
    \Phi_{A - B, i} (\vb{r})\: \vb{dr}, \\

  M_{p}^{(M)} & = &
  -\, \hbar\, (2\pi)^{3}\, \displaystyle\frac{\mu_{N}\,  m_{\rm p}}{\mu}\;
  \displaystyle\sum\limits_{\alpha=1,2}
  \displaystyle\int\limits_{}^{}
    \Phi_{A - B, f}^{*} (\vb{r})\;
  \vb{M}_{\rm eff} (\vb{k}_{\rm ph}, \vb{r}) \cdot
  e^{-i\, \vb{k}_{\rm ph} \vb{r}} \cdot
  \Bigl[ \vb{\displaystyle\frac{d}{dr}} \times \vb{e}^{(\alpha)} \Bigr]
  \Phi_{A - B, i} (\vb{r})\: \vb{dr},
\end{array}
\label{eq.13.1.2}
\end{equation}

\vspace{-5.5mm}
\begin{equation}
\begin{array}{lcl}
  M_{k} & = &
  i\, \hbar\, (2\pi)^{3}\,
  \mu_{N}\,
  \displaystyle\sum\limits_{\alpha=1,2}
    \bigl[ \vb{k_{\rm ph}} \cp \vb{e}^{(\alpha)} \bigr]
  \displaystyle\int\limits_{}^{}
    \Phi_{A - B, f}^{*} (\vb{r})\;
  \biggl\{
    e^{-i\, c_{B}\, \vb{k_{\rm ph}} \vb{r}}\, \vb{D}_{A,\, {\rm k}} +
    e^{ i\, c_{A}\, \vb{k_{\rm ph}} \vb{r}}\, \vb{D}_{B,\, {\rm k}}
  \biggr\} \cdot
  \Phi_{A - B, i} (\vb{r})\; \vb{dr},
\end{array}
\label{eq.13.1.3}
\end{equation}

\vspace{-5.5mm}
\begin{equation}
\begin{array}{lll}
\vspace{-1.0mm}
  M_{\Delta E} & = &
  -\, (2\pi)^{3}\,
  2\, \mu_{N}
  \displaystyle\sum\limits_{\alpha=1,2} \vb{e}^{(\alpha)}
  \displaystyle\int\limits_{}^{}
    \Phi_{A - B, f}^{*} (\vb{r})\; \times \\
  & \times &
  \biggl\{
    e^{-i\, c_{B}\, \vb{k_{\rm ph}} \vb{r}}\, \vb{D}_{A 1,\, {\rm el}} +
    e^{ i\, c_{A}\, \vb{k_{\rm ph}} \vb{r}}\, \vb{D}_{B 1,\, {\rm el}} -
    \displaystyle\frac{m_{\rm p}}{m_{A}}\, e^{-i\, c_{B}\, \vb{k_{\rm ph}} \vb{r}}\, \vb{D}_{A 2,\, {\rm el}} -
    \displaystyle\frac{m_{\rm p}}{m_{B}}\, e^{ i\, c_{A}\, \vb{k_{\rm ph}} \vb{r}}\, \vb{D}_{B 2,\, {\rm el}}
  \biggr\} \cdot
  \Phi_{A - B, i} (\vb{r})\; \vb{dr},
\end{array}
\label{eq.13.1.4}
\end{equation}

\vspace{-6.5mm}
\begin{equation}
\begin{array}{lll}
\vspace{-0.5mm}
  M_{\Delta M} & = &
  -\, i\, (2\pi)^{3}\,
  \mu_{N}\,
  \displaystyle\sum\limits_{\alpha=1,2}
  \displaystyle\int\limits_{}^{}
    \Phi_{A - B, f}^{*} (\vb{r})\;
  \biggl\{
    e^{-i\, c_{B}\, \vb{k_{\rm ph}} \vb{r}}\; D_{A 1,\, {\rm mag}} (\vb{e}^{(\alpha)}) +
    e^{ i\, c_{A}\, \vb{k_{\rm ph}} \vb{r}}\; D_{B 1,\, {\rm mag}} (\vb{e}^{(\alpha)})\; - \\
  & - &
    e^{-i\, c_{B}\, \vb{k_{\rm ph}} \vb{r}}\; D_{A 2,\, {\rm mag}} (\vb{e}^{(\alpha)}) -
    e^{ i\, c_{A}\, \vb{k_{\rm ph}} \vb{r}}\; D_{B 2,\, {\rm mag}} (\vb{e}^{(\alpha)})
  \biggr\} \cdot
  \Phi_{A - B, i} (\vb{r})\; \vb{dr}.
\end{array}
\label{eq.13.1.5}
\end{equation}
\end{widetext}

\vspace{-2.5mm}
\noindent
Here, $\vb{K}_{i}=\vb{K}_{f}+\vb{k}_{\rm ph}$, $c_{A}=m_{A}/(m_{A}+m_{B})$, $c_{B}=m_{B}/(m_{A}+m_{B})$, 
$m_{A}$ and $m_{B}$ are masses of two nuclei, and $m_{\rm p}$ is the proton mass. The reduced mass of the two nuclei is $\mu=m_{A}m_{B}/(m_{A}+m_{B})$, $\vb{k}_{\rm ph}$ is the photon wave vector, and $w_{\rm ph}=k_{\rm ph}c=|\vb{k}_{\rm ph}|c$. The function $\Phi_{A-B}(\vb{r})$ describes the relative motion of the two nuclei during the scattering. The quantities $F_{A,{\rm el}}$, $F_{B,{\rm el}}$, $\vb{F}_{A,{\rm mag}}$, $\vb{F}_{B,{\rm mag}}$, $D_{A,P{\rm el}}$, $D_{B,P{\rm el}}$, $\vb{F}_{A,P{\rm mag}}$, $\vb{F}_{B,P{\rm mag}}$, $\vb{D}_{A,{\rm k}}$, $\vb{D}_{B,{\rm k}}$, $\vb{D}_{A1,{\rm el}}$, $\vb{D}_{A2,{\rm el}}$, $\vb{D}_{B1,{\rm el}}$, $\vb{D}_{B2,{\rm el}}$, $D_{A1,{\rm mag}}$, $D_{A2,{\rm mag}}$, $D_{B1,{\rm mag}}$, and $D_{B2,{\rm mag}}$ are electric and magnetic form factors calculated in the Supplemental Material~\cite{thispaper.2026.Supplementary}. The effective electric charge $Z_{\rm eff}$ and effective magnetic moment $\vb{M}_{\rm eff}$ are given by [see Eq.~(S18) of the Supplemental Material~\cite{thispaper.2026.Supplementary}]

\vspace{-3.0mm}
\begin{equation}
\begin{array}{llllll}
\vspace{0.5mm}
  Z_{\rm eff} (\vb{k}_{\rm ph}, \vb{r}) & = & 
  e^{i\, \vb{k_{\rm ph}} \vb{r}}\,
  \Bigl[
    e^{-i\, c_{B} \vb{k_{\rm ph}} \vb{r}}\, \displaystyle\frac{m_{B}}{m_{A} + m_{B}}\, F_{A,\, {\rm el}} 
    (\vb{k_{ph}}) \\

  & - &
    e^{ i\, c_{A} \vb{k_{\rm ph}} \vb{r}}\, \displaystyle\frac{m_{A}}{m_{A} + m_{B}}\, 
    F_{B,\, {\rm el}} (\vb{k_{ph}})
  \Bigr], \\

  \textbf{M}_{\rm eff} (\vb{k}_{\rm ph}, \vb{r}) & = &
  \displaystyle\frac{e^{i\, \vb{k_{\rm ph}} \vb{r}}}{m_{\rm p}}
  \Bigl[
    e^{-i\, c_{B} \vb{k_{\rm ph}} \vb{r}}\, \displaystyle\frac{m_{B}\, m_{A}\, \vb{F}_{A,\, {\rm mag}} (\vb{k_{ph}})}{m_{A} + m_{B}}\, \\

  & - &
    e^{ i\, c_{A} \vb{k_{\rm ph}} \vb{r}}\, \displaystyle\frac{m_{A}}{m_{A} + m_{B}}\, m_{B}\, 
    \vb{F}_{B,\, {\rm mag}}  (\vb{k_{ph}})
  \Bigr].
\end{array}
\label{eq.13.1.6}
\end{equation}

\vspace{-2.0mm}
\noindent 
In previous bremsstrahlung calculations for proton--nucleus scattering, the monopole approximation to $Z_{\rm eff}(\vb{k}_{\rm ph},\vb{r})$ was used, namely $Z_{\rm eff}^{\rm (mon,0)}=(Z_A m_B-Z_B m_A)/(m_A+m_B)$ in the limit $\vb{k}_{\rm ph}\to0$ [see Eq.~(25) of Ref.~\cite{Maydanyuk.2023.PRC.delta},
$Z_A = F_{A, {\rm el}} (\vb{k_{ph}})$, 
$Z_B = F_{B, {\rm el}} (\vb{k_{ph}})$
at this limit]. For the symmetric reaction $\isotope[124]{Sn}+\isotope[124]{Sn}$, however, this approximation gives $Z_{\rm eff}^{\rm (mon,0)}=0$. It would therefore remove the coherent contribution in Eqs.~(\ref{eq.13.1.2}) and (\ref{eq.13.1.3}), which is an unphysical consequence of an over-simplified effective-charge approximation. We therefore use the more accurate approximation $Z_{\rm eff}(\vb{k}_{\rm ph},\vb{r})\approx Z_{\rm eff}^{\rm (mon,0)}(k_{\rm ph})=-i z_A\sin(c_A k_{\rm ph}r)$ and obtain
[see Eq.~(S26) in 
\cite{thispaper.2026.Supplementary}, for details]

\vspace{-2.0mm}
\noindent 
\begin{equation}
\begin{array}{llllll}
\vspace{0.5mm}
M_{p}^{(E)} =
  -\, \hbar\, (2\pi)^{3}\,
  \displaystyle\frac{\mu_{N}}{\sqrt{3}}\,
  \displaystyle\frac{m_{\rm p}}{\mu}\, \\

\vspace{1.5mm}
  \hspace{10.0mm} \times\; 
  \Bigl(
    J_{1}^{\rm (sym)} (0,1,0) -
    \displaystyle\frac{47}{40} \sqrt{\displaystyle\frac{1}{2}}\, J_{1}^{\rm (sym)} (0,1,2)
  \Bigr), \\

\vspace{0.5mm}
M_{p}^{(E)} + M_{p}^{(M)} =
  M_{p}^{(E)}
  \Bigl( 1 - i \displaystyle\frac{m_{\rm p}}{2\, \mu} \bar{\mu}_{\rm pn}^{(A)} \Bigr), \\

\vspace{1.5mm}
  M_{\Delta M} =
  \hbar (2\pi)^{3} \sqrt{3}
  \mu_{N}\, k_{\rm ph}
  \Bigl\{
    \displaystyle\frac{A-1}{2A} \bar{\mu}_{\rm pn}^{\rm (A)}
      Z_{\rm A} (\vb{k}_{\rm ph}) 
      \tilde{J} (c_{A}, 0,1,0)
  \Bigr\}, \\

  M_{k} =
  - \hbar (2\pi)^{3} \sqrt{3}
  \mu_{N} k_{\rm ph}
  \Bigl\{ \bar{\mu}_{\rm pn}^{(A)} \, Z_{\rm A} (\vb{k}_{\rm ph}) \, \tilde{J}\, (c_{A}, 0,1,0) \Bigr\}.
\end{array}
\label{eq.multimple.4}
\end{equation}

\vspace{-2.0mm}
\noindent 
In these expressions, $\bar{\mu}_{\rm pn}^{\rm (A)}=\mu_{\rm p}+\kappa_{A}\mu_{\rm n}$, $\bar{\mu}_{\rm pn}^{\rm (B)}=\mu_{\rm p}+\kappa_{B}\mu_{\rm n}$, $\kappa_{A}=N_A/Z_A$, and $\kappa_{B}=N_B/Z_B$. Here, $Z_A$ and $N_A$ ($Z_B$ and $N_B$) are the numbers of protons and neutrons in nucleus $A$ ($B$), while $\mu_{\rm p}$ and $\mu_{\rm n}$ are the magnetic moments of the proton and neutron. For $\isotope[124]{Sn}+\isotope[124]{Sn}$, $\bar{\mu}_{\rm pn}^{\rm (A)}=\bar{\mu}_{\rm pn}^{\rm (B)}=-0.038\:455\:895\:78\,\mu_N$. The radial integrals are

\vspace{-1.5mm}
\begin{equation}
\begin{array}{llllll}
\vspace{1.0mm}
  J_{1}^{\rm (sym)}(l_{i},l_{f},n) & = &
  \displaystyle\int\limits^{+\infty}_{0} 
    Z_{\rm eff}^{\rm (mon,\, 0)} (k_{\rm ph})\;
    \displaystyle\frac{dR_{i}(r, l_{i})}{dr}\: \\
    
\vspace{1.0mm}
  & \times & \; 
    R^{*}_{f}(l_{f},r)\, j_{n}(k_{\rm ph}r)\; r^{2} dr, \\

  \tilde{J}\,(c, l_{i},l_{f},n) & = & \displaystyle\int\limits^{+\infty}_{0} R_{i}(l_{i}, r)\, R^{*}_{f}(l_{f},r)\, j_{n}(c\, k_{\rm ph}r)\; r^{2} dr.

\end{array}
\label{eq.multimple.5}
\end{equation}

\vspace{-1.0mm}
\noindent
Here, $R_{i,f}$ is the radial part of the wave function $\Phi_{A-B}(\vb{r})$, which describes the relative motion of the two nuclei, while the internal relative motion of the nucleons inside each nucleus is not treated explicitly. The indices $i$ and $f$ refer to the states before and after photon emission, respectively, and $j_n(k_{\rm ph}r)$ is the spherical Bessel function of order $n$.

Experimentally, the probability of bremsstrahlung photon emission per collision, $dM/dw_{\rm ph}$, was measured. In the model, we define the bremsstrahlung cross section following the formalism for proton--nucleus scattering~\cite{Maydanyuk.2023.PRC.delta} [see Eq.~(29) therein] and calculate the bremsstrahlung probability by renormalizing the cross section as
\begin{equation}
\begin{array}{llllll}
  \displaystyle\frac{d^{2} P}{dw_{\rm ph}\, d\, \Omega_{\rm ph}} =
  N \cdot \displaystyle\frac{d \sigma}{dw_{\rm ph}}, &

  \displaystyle\frac{d \sigma}{dw_{\rm ph}} =
    \displaystyle\frac{w_{\rm ph}\,E_{i}}{2\pi\,c^{5}\,k_{i}}\,
    \bigl| M_{\rm full} \bigr|^{2},

\end{array}
\label{eq.multimple.6}
\end{equation}
where $N$ is a single global normalization factor found concerning to the full energy region of the measured data. 
This factor used in calculation of the coherent and incoherent contributions.
Incoherent-to-coherent ratio is not changed at variations of this factor.
The matrix elements are calculated using wave functions with quantum numbers $l_i=0$, $l_f=1$, and $l_{\rm ph}=1$. Here, $l_i$ and $l_f$ are the orbital quantum numbers of $\Phi_{A-B}(\vb{r})$ before and after photon emission, respectively, and $l_{\rm ph}$ is the photon orbital quantum number in the multipole expansion.

The wave functions 
are calculated concerning energy of the relative motion between the two nuclei at the center-of-mass frame,
$E_{\rm scm}=\mu E_{B,{\rm lab}}/m_B=1550$~MeV, where $E_{B,{\rm lab}}=124\times25$~MeV $=3100$~MeV is the kinetic energy of the $\isotope[124]{Sn}$ beam in the laboratory frame
(see Eqs.~(S32) in~\cite{thispaper.2026.Supplementary}).

\begin{figure}[htbp]
\centerline{\includegraphics[width=90mm]{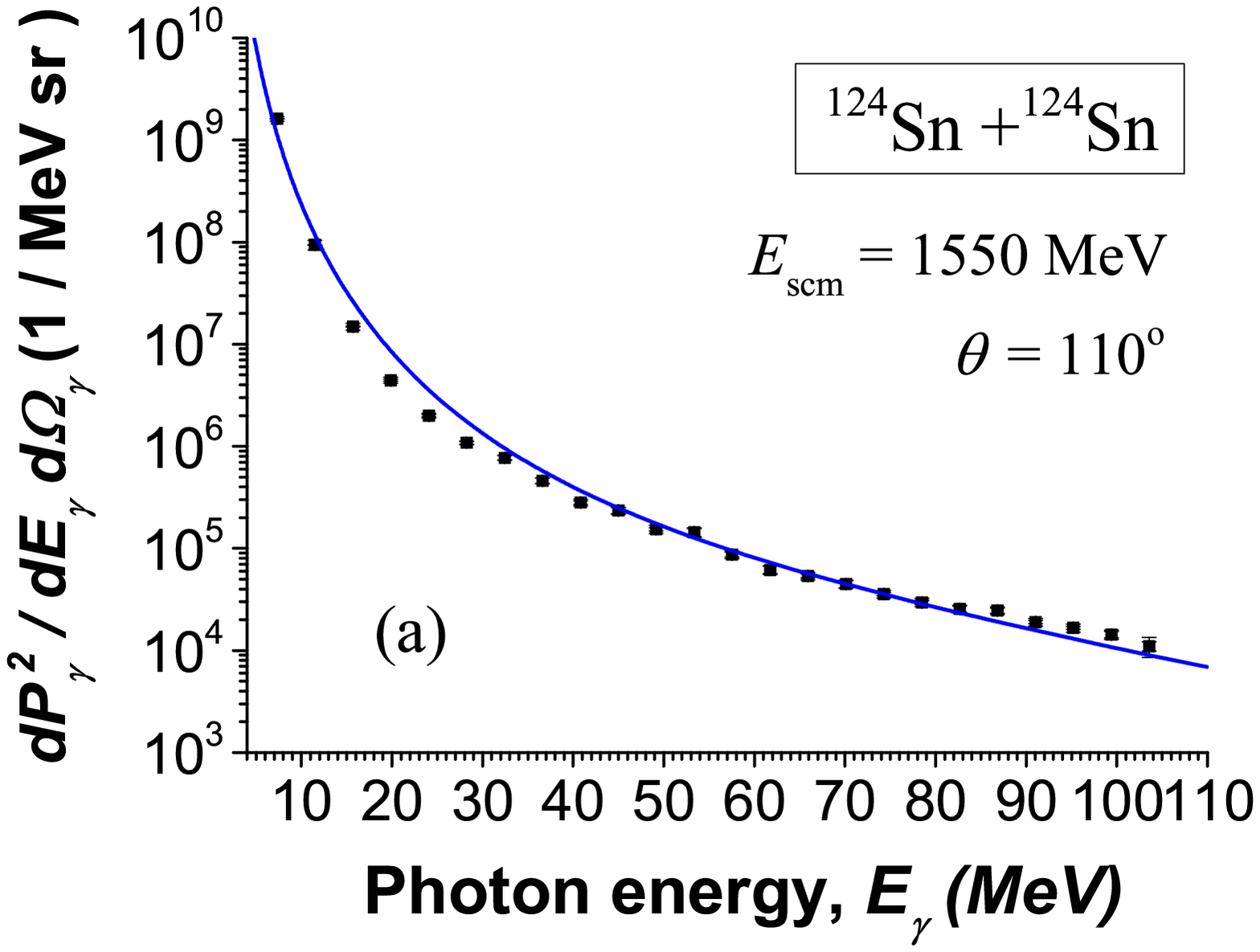}}
\vspace{-1.0mm}
\centerline{\includegraphics[width=90mm]{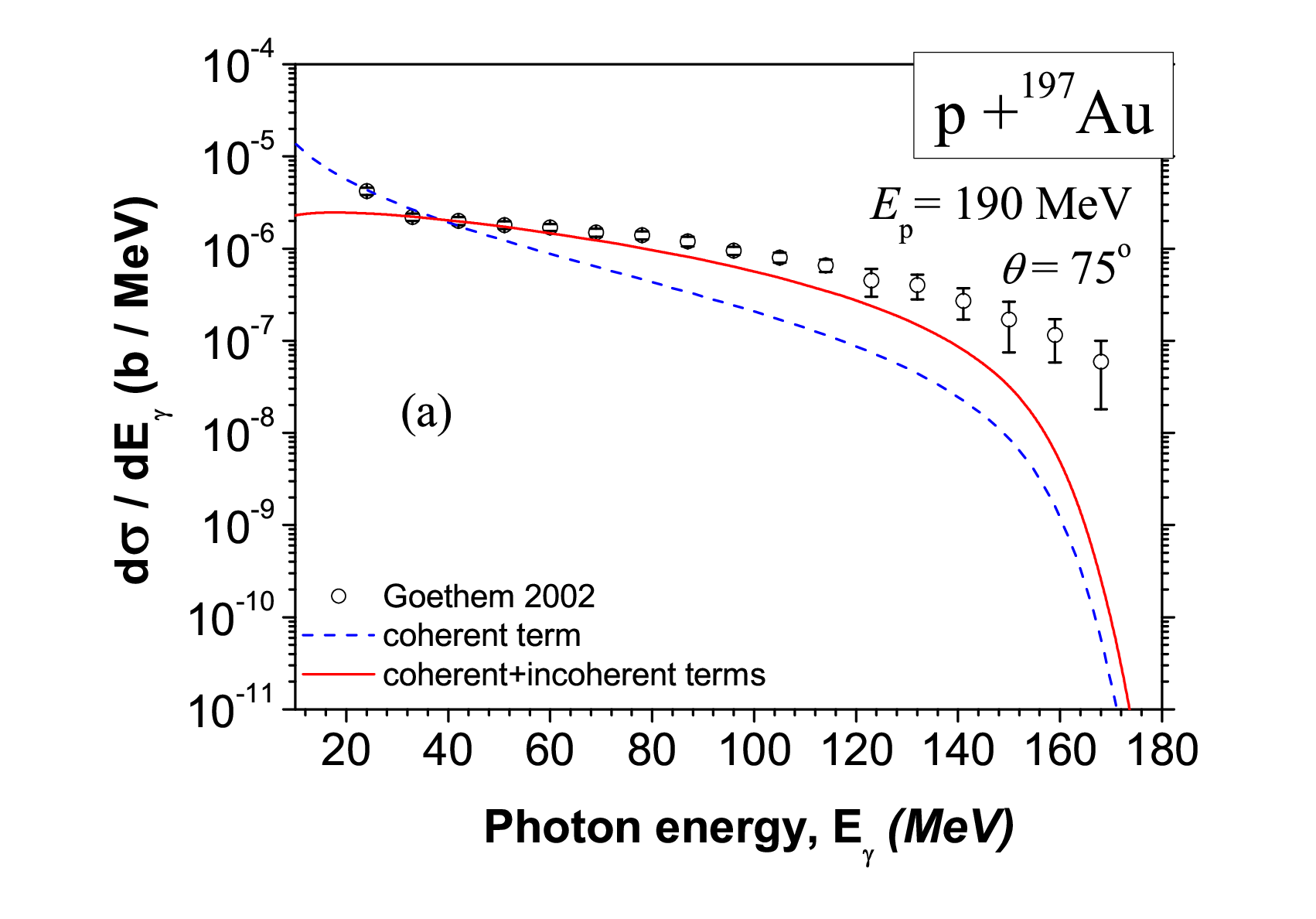}}
\vspace{-4mm}
\caption{\small (Color online)
Panel (a): 
Calculated full spectrum of bremsstrahlung photons for $\isotope[124]{Sn}+\isotope[124]{Sn}$ at 
$25\:MeV/u$ 
compared with experimental data~\cite{br124Sn124Snexp.2026.PRC}. The bremsstrahlung probability is calculated using Eq.~(\ref{eq.multimple.6}). The full matrix element includes the coherent terms 
$M_{p}^{(E)}$ and $M_{p}^{(M)}$ and the incoherent terms $M_{\Delta M}$ and $M_k$ defined in Eq.~(\ref{eq.multimple.4}); the radial integrals are defined in Eq.~(\ref{eq.multimple.5}). The photon-emission angle is $\theta=110^{\circ}\pm5^{\circ}$ with respect to the beam direction, and the calculated full spectrum is normalized to the experimental data. 
Panel (b): 
Bremsstrahlung photons in 
$p+\isotope[197]{Au}$ at a proton beam energy of 190~MeV~\cite{Goethem.2002.PRL,Maydanyuk_Zhang.2015.PRC,Maydanyuk.2023.PRC.delta}.
\label{fig.1}}
\end{figure}

\vspace{1.5mm}
\emph{Analysis.}\;
A new measurement of bremsstrahlung $\gamma$ rays was performed for the scattering of $\isotope[124]{Sn}$ nuclei on a $\isotope[124]{Sn}$ target using the Compact Spectrometer for Heavy IoN Experiment (CSHINE)~\cite{br124Sn124Snexp.2026.PRC,Xu:2025prr}. This spectrometer has been constructed, maintained, and continuously upgraded at the Radioactive Ion Beam Line in Lanzhou (RIBLL-1)~\cite{Wang.2021.NuclSciTech,Wang.2021.NuclInstMeth}. A $\isotope[124]{Sn}$ beam at 25~MeV/nucleon was delivered by the Heavy Ion Research Facility in Lanzhou (HIRFL) and directed onto a $\isotope[124]{Sn}$ target installed in the chamber located at the final focal plane of RIBLL-1. The measured spectrum was analyzed using IBUU calculations that incorporate the photon-production channel from the $np$ process~\cite{br124Sn124Snexp.2026.PRC,Xu:2025prr}. The data are shown in Fig.~\ref{fig.1}.

We applied the model described above to bremsstrahlung emission in this reaction. The full calculated spectrum, including both coherent and incoherent contributions, is compared with the experimental data in Fig.~\ref{fig.1}(a).
The calculated spectrum agrees well with the experimental data over the full measured energy range. This agreement provides an important validation of the model and supports its use for estimating the individual coherent and incoherent contributions.

Bremsstrahlung emission in proton--nucleus scattering has been extensively studied both experimentally and theoretically~\cite{Goethem.2002.PRL,Maydanyuk_Zhang.2015.PRC,Maydanyuk.2023.PRC.delta}. The most accurate study of this type is for $p+\isotope[197]{Au}$, shown in Fig.~\ref{fig.1}(b). The comparison with Fig.~\ref{fig.1}(a) reveals a qualitative difference between proton--nucleus and nucleus--nucleus scattering.  
In proton--nucleus scattering, the spectrum contains a pronounced hump, whereas the $\isotope[124]{Sn}+\isotope[124]{Sn}$ spectrum decreases monotonically and has a nearly logarithmic shape. 
The joint unified model describes these opposite spectral behaviors with good accuracy.

The dominance of incoherent emission in proton--nucleus scattering was established in earlier studies~\cite{Goethem.2002.PRL,Maydanyuk_Zhang.2015.PRC,Maydanyuk.2023.PRC.delta}. In particular, this conclusion follows from analyses of $p+\isotope[208]{Pb}$ at proton beam energies of 140 and 145~MeV and of $p+\isotope[12]{C}$, $p+\isotope[58]{Ni}$, $p+\isotope[107]{Ag}$, and $p+\isotope[197]{Au}$ at $E_{\rm p}=190$~MeV. 
Last estimates gave an incoherent-to-coherent ratio of about 
$10^{3}$--$10^{5}$~\cite{Maydanyuk.2023.PRC.delta}. A characteristic manifestation of incoherent emission is the hump in the photon spectrum. In proton--nucleus scattering, this hump is clearly visible in the middle part of the measured spectrum [see Fig.~\ref{fig.1}(b) and Fig.~2 of Ref.~\cite{Goethem.2002.PRL}]. Its dependence on the magnitude of the incoherent contribution is also shown in Fig.~5 for $p+\isotope[208]{Pb}$ at $E_{\rm p}=145$~MeV, Fig.~6(a) for $p+\isotope[12]{C}$ at $E_{\rm p}=190$~MeV, Fig.~6(b) for $p+\isotope[197]{Au}$ at $E_{\rm p}=190$~MeV, Fig.~7(a) for $p+\isotope[58]{Ni}$ at $E_{\rm p}=190$~MeV, and Fig.~7(b) for $p+\isotope[107]{Ag}$ at $E_{\rm p}=190$~MeV in Ref.~\cite{Maydanyuk_Zhang.2015.PRC}. By contrast, the measured $\isotope[124]{Sn}+\isotope[124]{Sn}$ spectrum in Fig.~\ref{fig.1}(a) shows no such hump, indicating that the incoherent contribution is small and that coherent emission dominates.

\begin{figure}[htbp]
\centerline{\includegraphics[width=90mm]{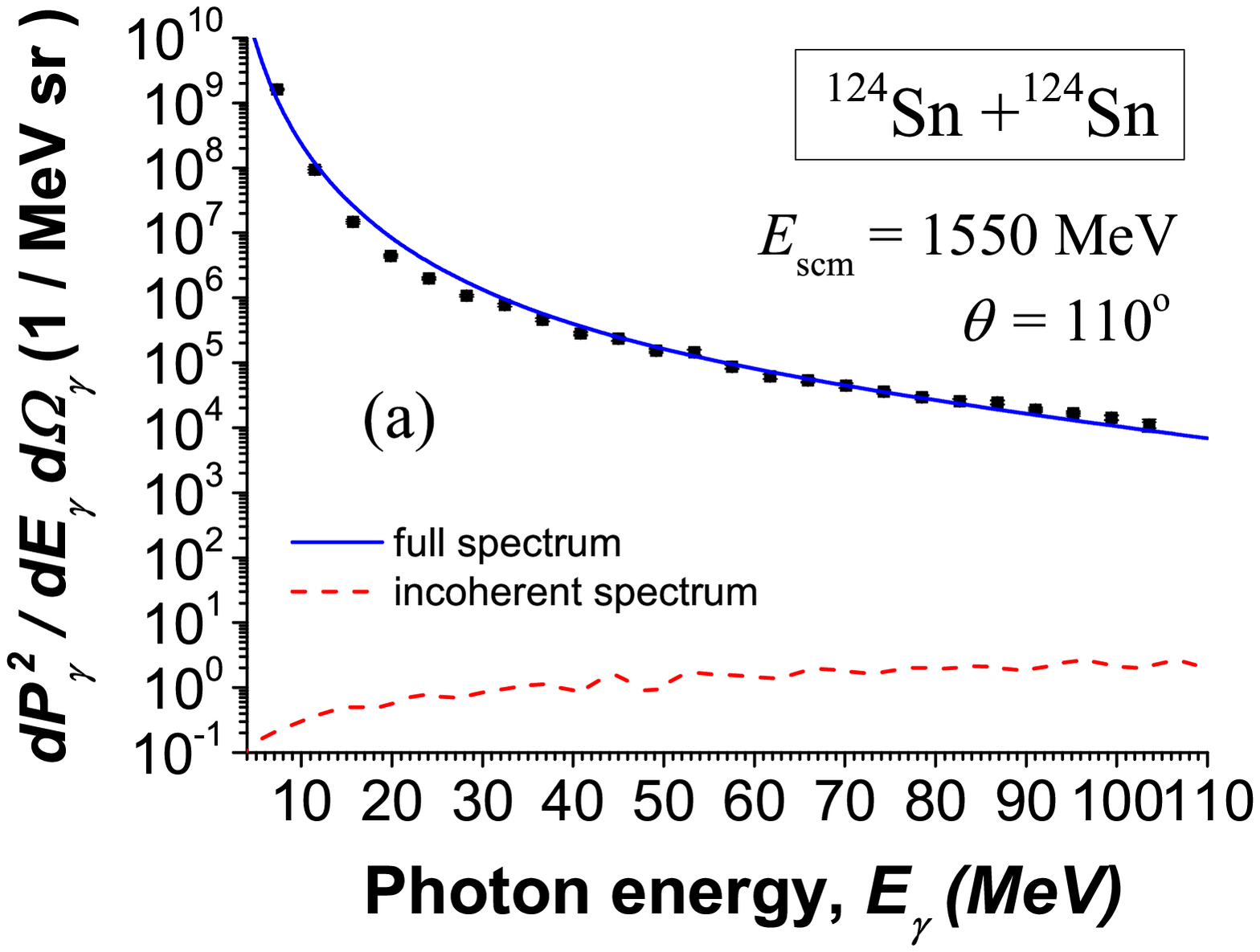}}
\centerline{\includegraphics[width=90mm]{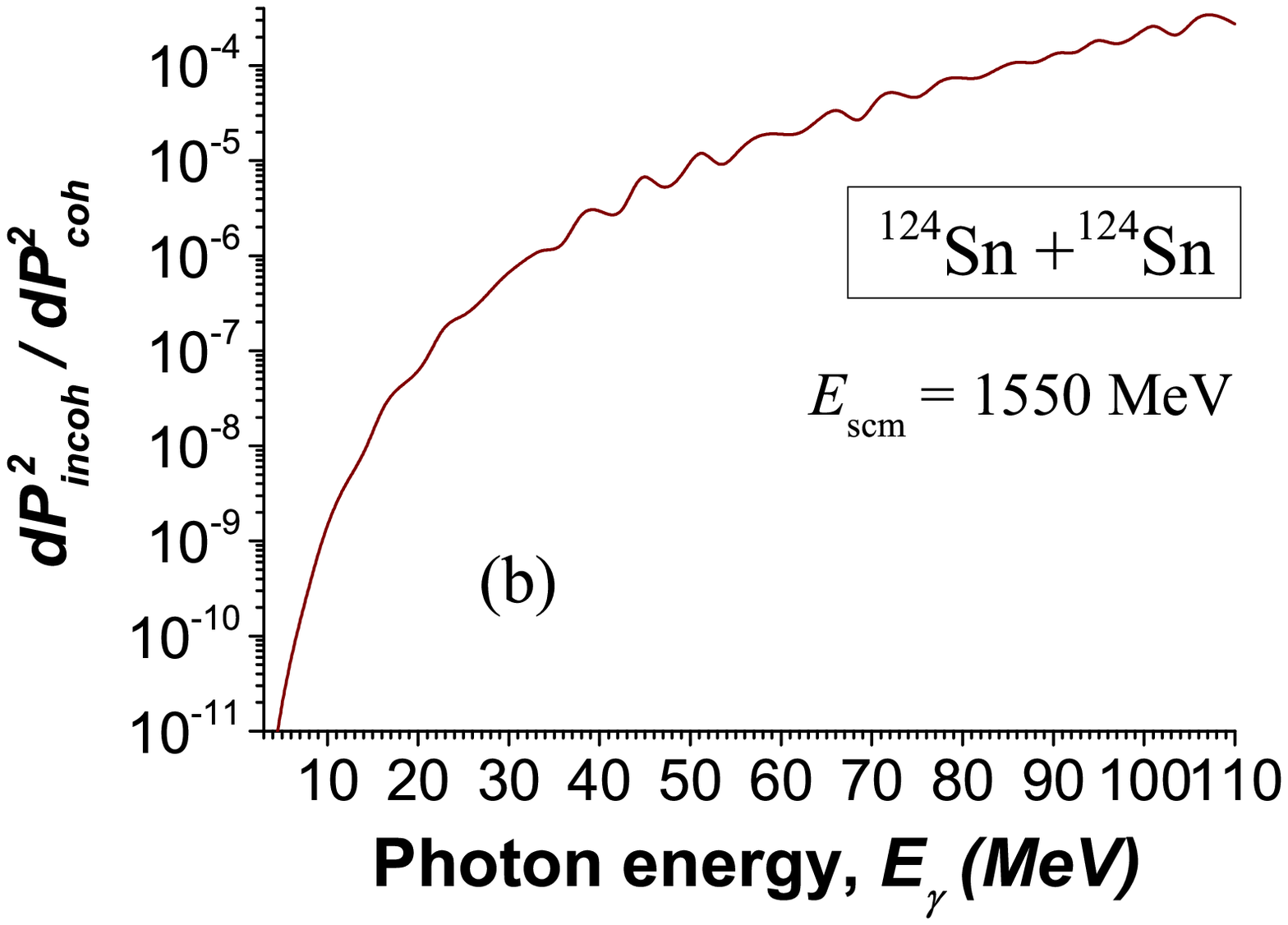}}
\vspace{-2.0mm}
\vspace{-4mm}
\caption{\small (Color online)
Panel (a): Incoherent contribution and full bremsstrahlung spectrum for $\isotope[124]{Sn}+\isotope[124]{Sn}$ scattering at a beam energy of $25\:MeV/u$ (kinetic energy of the relative motion in the center-of-mass frame, $E_{\rm scm}=1550$~MeV), compared with experimental data~\cite{br124Sn124Snexp.2026.PRC}. 
Panel (b): 
Ratio between incoherent and coherent contributions shown in Fig.~2(a)
as a function of photon energy. The incoherent contribution is many orders of magnitude smaller than the coherent contribution. This behavior is opposite to that observed in proton--nucleus scattering, where incoherent bremsstrahlung dominates~\cite{Goethem.2002.PRL,Maydanyuk.2023.PRC.delta,Maydanyuk_Zhang.2015.PRC}.
\label{fig.2}}
\end{figure}
To quantify this phenomenon, we calculated the incoherent and coherent contributions, as shown in Fig.~\ref{fig.2}(a), relative to the full spectrum. The coherent bremsstrahlung component clearly dominates in this reaction. This dominance is even more evident from the incoherent-to-coherent ratio shown in Fig.~\ref{fig.2}(b), especially at low photon energies.

\emph{Conclusions.}\;
We have presented a combined experimental and theoretical study of bremsstrahlung emission in $\isotope[124]{Sn}+\isotope[124]{Sn}$ scattering at a beam energy of $E=25~MeV/u$. We extended the established bremsstrahlung model for proton--nucleus scattering~\cite{Maydanyuk_Zhang.2015.PRC,Maydanyuk.2023.PRC.delta,Maydanyuk.Tsushima.Ramalho.PMZhang.2026.PLB} by developing a formalism for nucleus--nucleus scattering and a unified treatment of coherent and incoherent emission in such reactions. The resulting model describes the measured spectrum with high accuracy over the full experimental energy range [see Fig.~\ref{fig.1}(a)].

So, at first time, we establish the qualitative difference between the two spectra:
(1) In proton--nucleus scattering, the spectrum contains a pronounced hump
[see Fig.~\ref{fig.1}(b)],
(2) In $\isotope[124]{Sn} + \isotope[124]{Sn}$ scattering, the spectrum decreases monotonically and has a nearly logarithmic shape [see Fig.~\ref{fig.1}(a)].

The calculations demonstrate the dominant role of coherent bremsstrahlung emission in $\isotope[124]{Sn}+\isotope[124]{Sn}$ scattering [see Fig.~\ref{fig.2}]. We find that the electric charges of the nucleons play a much larger role in bremsstrahlung emission than their magnetic moments. In particular, the dominant coherent electric matrix element $M_{p}^{(E)}$ depends on the electric charges of the protons in the nuclei through the effective electric charge 
$Z_{\rm eff}^{\rm (mon,0)}(k_{\rm ph})$. 
The term $M_{p}^{(M)}$, by contrast, contains the magnetic moments of the nucleons and is much smaller: from Eq.~(\ref{eq.multimple.4}), the calculation gives $m_{\rm p}\bar{\mu}_{\rm pn}^{\rm (A)}/(2\mu)=0.000\:310\:128$.

The situation is opposite in proton--nucleus scattering, where the magnetic moments of the nucleons are much more important than their electric charges for bremsstrahlung emission. Recent estimates for $p+\isotope[197]{Au}$ at $E_{\rm p}=190$~MeV give an incoherent-to-coherent ratio of 
$10^{3}$--$10^{5}$~\cite{Maydanyuk.2023.PRC.delta,Goethem.2002.PRL}. The agreement between the measured heavy-ion data and the present calculations confirms, with high precision over the studied energy range, that $\isotope[124]{Sn}+\isotope[124]{Sn}$ scattering realizes the opposite regime, in which coherent emission dominates.

\emph{Acknowledgements.}\:
S.P.M. thanks Sun Yat-Sen University for its warm hospitality and support. 
The authors thank 
Prof. Xiao Zhigang for the numerous discussions, help in understanding and analysis of physics and results,
Prof.~V.~S.~Vasilevsky for useful recommendations on several aspects of nuclear-collision physics, Prof.~A.~G.~Magner for valuable discussions of compound nuclear systems and fusion, and Prof.~G.~Wolf for fruitful discussions of transport models and high-energy heavy-ion collisions. 
This work was supported by 
National Key R\&D 
Program of China 
(grant No.~2024YFE0109802), 
the National Natural Science Foundation of China (grants Nos. 12175320, 12375084), 
the Natural Science Foundation of Guangdong Province, China (Grant No. 2022A1515010280).

\bibliography{reference}

\end{document}